\documentclass[11pt,a4paper]{article}
\usepackage{eepic}
\usepackage{epsf}
\usepackage{amsmath}
\usepackage{graphicx}
\usepackage[normalem]{ulem}
\usepackage{hyperref}
\usepackage[superscript]{cite}
\newcommand \bea {\begin{eqnarray}}
\newcommand \eea {\end{eqnarray}}
\usepackage{placeins}
\begin{document}
\baselineskip=12pt
\begin{center}
{\LARGE{Computational investigation on non-linear optical properties of 
hexaphyrin and core modified hexaphyrins}}\\
\vspace*{1.0cm}
Sumit Naskar$^a$, Mousumi Das$^{*a,b}$ \\
$^a${\it Department of Chemical Sciences, Indian Institute of Science
Education and Research Kolkata}
\\
$^b${\it Centre for Advanced Functional Materials (CAFM), Indian Institute of Science Education and Research Kolkata,
Mohanpur - 741246, India, email: mousumi@iiserkol.ac.in}\\
\end{center}

\vspace*{2.0cm}
\begin{abstract}
Expanded porphyrin-based (Hexaphyrins) sensitizers are promising due to their excellent
light harvesting feature in dye-sensitized solar cell (DSSC). We
calculated the low-lying excitations of expanded porphyrins (EPs) as hexaphyrin and core modified hexaphyrin structures 
using Time-Dependent Density Functional Theory. Our calculation showed the EPs (both hexaphyrin and core modified hexaphyrin)
have broad range of absorption band suitable for harvesting the visible and near infrared region of solar spectrum. 
All EPs studied here satisfy the energy condition of singlet fission (SF). SF is the process in which the theoretical
limit of Shockley-Quiesser (SQ) (33\%) can be overcome in single junction solar cell. The non-linear optical properties
like first hyper polarizability $\beta$ and second order hyper polarizability $\gamma$ were calculated using coupled 
perturbed Hartree-Fock approach. From the second order NLO properties we carried out degenerate four wave mixing (DFWM)
component ($\gamma^{(2)}(-\omega;\omega,\omega,-\omega$)) and finally quadratic non linear refractive indices of these
EPs are calculated.
Calculation showed EPs are promising as organic dye for the opto-electronic applications and useful for high efficiency 
DSSC and also useful for potential NLO materials as 
their hyper polarizabilities showed higher order non linearities.
\end{abstract}
\newpage
\section{INTRODUCTION}
The organic conjugated 
molecules with diverse $\pi$ electronic cloud delocalised all over 
are capable
of showing various linear and non-linear optical (NLO) responses. 
Energy ordering of low-lying excited states of conjugated organic molecules show promise in the field of singlet fission (SF).\cite{smith2010singlet,naskar2020use}
SF is a multiexciton generation process in which from one excited singlet exciton two triplet excitons are generated. The
generation of multiple excitons at a time overcomes the Shockley-Queisser (SQ) theoretical limit ($\sim$ 33\%) of photoconversion efficiency in single junction solar-cells\cite{shockley1961detailed}. The energy criteria for SF is developed by Paci $et\ al.$ in 2006 as:\cite{paci2006SF}
\begin{equation}
2E(T_1-S_0)\leq E(S_1-S_0)
\label{SF1}
\end{equation}
and
\begin{equation}
2E(T_1-S_0)< E(T_2-S_0)
\label{SF2}
\end{equation}
The first condition \ref{SF1} provides the essential energy condition for the formation of two triplet excitons out of one singlet exciton. The second condition \ref{SF2} nullify the triplet-triplet annihilation (TTA).\\
The non-linear optical response to applied electric fields is described
in terms of the hyperpolarizabilities of the molecules.
The effort was made earlier
to calculate the first
order hyperpolarizability in organic polar molecules 
due to their application in modulation of 
electro-optic effect, sensing, imaging, microfabrication and 
many more\cite{oudar-chemla,marder}. 
The higher order nonlinearity was measured using
$Z$ scan method in crystals and also in molecules\cite{bache2013anisotropic,makhal2016third}. 
Second and third order dynamic nonlinearity
was theoretically observed recently in several acid derivatives\cite{karakas2019theoretical}.\\
In recent times porphyrin and metalloporphyrins are subject 
of immense interest due to their large non-linear responses such as 
third harmonic generation (THG) and two-photon absorption (TPA) cross sections
\cite{nalwa-jpcA-1993},\cite{drobizhev-cpl-2002}. In this regard, expanded form
of porphyrins (EPs) attract a considerable attention in recent times 
due to their
interesting linear and nonlinear optical responses\cite{simil-JPCA-2013}.  
Unlike porphyrin, EPs have more than four pyrrole rings or meso-links. It is
also noted that number of pyrrole rings and meso-links in EPs guides their absorption spectra
covering the entire visible range. EPs may have twisted geometry and possess 4n as well as 4n+2
$\pi$-electrons. Theoretical study based on correlated model
shows EPs possessing 4n+2 electrons have large two-photon 
absorption than that of 4n electrons.
\cite{simil-JPCA-2013}

In this paper,
hexaphyrin, 
core modified hexaphyrin 
such as Ni-hexaphyrin, 
Cu-hexaphyrin, 
and Au-hexaphyrin 
are considered to calculate linear and nonlinear optical properties.
In modified hexaphyrin structures, the metals at the center in EPs
provide a perturbation to the systems.
The energy ordering the low-lying excited states in singlet and triplet
subspaces are reported. 
Our calculation showed that 
all the EPs satisfy the first condition of SF\ref{SF1},however the 
second condition of SF\ref{SF2} is not satisfied. If one can tune the TTA cross section judicially then these material
can be used as potential SF materials. 
\\The non-linear optical properties of these systems 
both in presence of frequency
($\omega$) and absence of it. 
From second order hyperpolarizability the quadratic nonlinear 
refractive index $n_2$
of these molecules are calculated which may help to 
modulate the frequency control over 
LASER emission\cite{ganeev2003nonlinear}. 
The sign of $n_2$ helps 
to identify the one or two photon 
frequency dominance.  
If the 
sign of $n_2$ is positive, two photon frequency 
dominates over one photon frequency and vice versa.
\begin{figure}
\begin{center}
\includegraphics[width=0.6\linewidth]{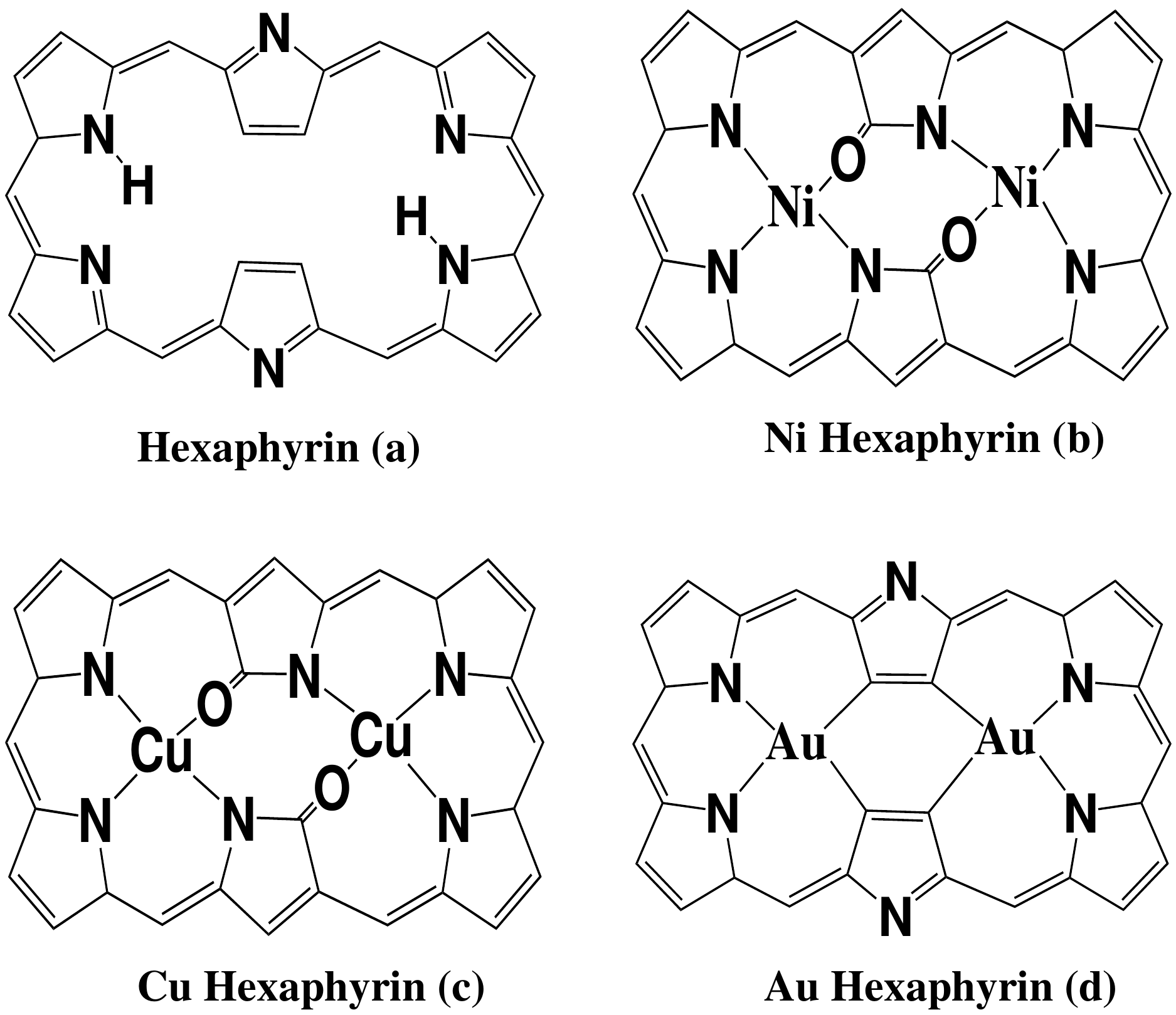}
\caption{\bf{Hexaphyrin($\bf a$) and core modified hexaphyrin 
systems($\bf b - d$) studied}}
\label{fig-hexaphyrin}
\end{center}
\end{figure}
\section{COMPUTATIONAL APPROACH}
These molecules hexaphyrin
($\bf a$), 
core modified hexaphyrin
such as Ni-hexaphyrin
($\bf b$), 
Cu-hexaphyrin
($\bf c$),  
and Au-hexaphyrin ($\bf c$) shown in Fig. \ref{fig-hexaphyrin} 
are first optimized 
in $Gaussian 09$ \cite{ref16} based on density functional theory (DFT) with B3LYP exchange correlation
functional.\cite{ref18,ref19}
The basis set used in optimization 
for C, H, N and O atoms is 6-31G(d,p) 
while LANL2DZ basis                              
set is used for metal atoms like Ni, Cu and Au.  
The low-lying excited states in singlet and triplet subspaces
are calculated on these optimized molecular 
structures($\bf a-d$) through Time Dependent Density Functional Theory (TDDFT) 
\cite{ref17}. The polarizabilities of a molecule is defined as
a response to the applied electric field ($\overrightarrow\epsilon$)
which induces a dipole moment($\overrightarrow\mu$). The perturbed Hamiltonian
($H^\prime$) due to electric field is defined as 
$\overrightarrow{H^\prime}=\overrightarrow\mu.\overrightarrow\epsilon$. 
The total energy modified due to such perturbation can be written in
Taylor's series as:
\begin{eqnarray}
E(\overrightarrow\epsilon)=E(0)+\sum\limits_{i}{}\left(\frac{\partial{E}}{\partial{\epsilon_{i}}}\right)\epsilon_{i}+\frac{1}{2}\sum\limits_{i}{}\sum\limits_{j}{}\left(\frac{\partial^{2}{E}}{\partial{\epsilon_{i}}{\partial{\epsilon_{j}}}}\right)\epsilon_{i}\epsilon_{j}\nonumber\\
+\frac{1}{6}\sum\limits_{i}{}\sum\limits_{j}{}\sum\limits_{k}{}\left(\frac{\partial^{3}{E}}{\partial{\epsilon_{i}}{\partial{\epsilon_{j}}}{\partial{\epsilon_{k}}}}\right)\epsilon_{i}\epsilon_{j}\epsilon_{k}+\frac{1}{24}\sum\limits_{i}{}\sum\limits_{j}{}\sum\limits_{k}{}\sum\limits_{l}{}\nonumber\\\times\left(\frac{\partial^{4}{E}}{\partial{\epsilon_{i}}{\partial{\epsilon_{j}}}{\partial{\epsilon_{k}}}{\partial{\epsilon_{l}}}}\right)\epsilon_{i}\epsilon_{j}\epsilon_{k}\epsilon_{l}+\hdots
\label{static-eq}
\end{eqnarray} 

The second term in Eqn. \ref{static-eq} on the
R.H.S determines the $i^{th}$ 
component of dipole moment and defined as:
\begin{equation}
\mu_{i}=-\left(\frac{\partial{E}}{\partial{\epsilon_{i}}}\right)
\end{equation}
The higher order terms in the R.H.S will give linear 
polarizability $\alpha_{ij}$, 
first hyper polarizability $\beta^{(1)}_{ijk}$ and second hyper 
polarizability $\gamma^{(2)}_{ijkl}$ defined as:
\begin{equation}
\alpha_{ij}=-\left(\frac{\partial^{2}{E}}{\partial{\epsilon_{i}}{\partial{\epsilon_{j}}}}\right)
\end{equation}
\begin{equation}
\beta_{ijk}=-\left(\frac{\partial^{3}{E}}{\partial{\epsilon_{i}}{\partial{\epsilon_{j}}}{\partial{\epsilon_{k}}}}\right)
\end{equation}
and 
\begin{equation}
\gamma^{(2)}_{ijkl}=-\left(\frac{\partial^{4}{E}}{\partial{\epsilon_{i}}{\partial{\epsilon_{j}}}{\partial{\epsilon_{k}}}{\partial{\epsilon_{l}}}}\right)
\end{equation}
The isotropic average of polarizability given by $\alpha_{avg}=\frac{1}{3}(\alpha_{XX}+\alpha_{YY}+\alpha_{ZZ})$, 
First hyperpolarizability $\beta$ and second hyperpolarizability $\gamma$ calculated in static condition using finite
electric field of 0.001 a.u. .\\
In presence of a time-dependent electric field 
$E=E_0+E_\omega cos(\omega t)$, 
the above Eqn. \ref{static-eq} can be rewritten as 
\cite{tarazkar2015theoretical}:
\begin{eqnarray}
E(\overrightarrow\epsilon)=E(0)+\alpha_{ij}(0;0)\epsilon_{0\mu} +\alpha_{ij}(-\omega;\omega)\epsilon_{\omega j} cos(\omega t)
+\frac{1}{2} \beta_{ijk}(0;0,0)\epsilon_{0 j}\epsilon_{0 k}\nonumber\\
 +\frac{1}{4} \beta_{ijk}(0;\omega,-\omega) 
\epsilon_{\omega j}\epsilon_{\omega k} +\beta_{ijk}(-\omega;0,\omega)
\epsilon_{0 j} \epsilon_{\omega k} cos(\omega t)\nonumber\\~
+\frac{1}{4}\beta_{ijk}(-2\omega;\omega,\omega)
+\frac{1}{6}\gamma^{(2)}_{ijkl}(0;0,0,0)\epsilon_{0 j}\epsilon_{0 k}\epsilon_{0 l}\nonumber\\
 +\frac{1}{2}\gamma^{(2)}_{ijkl}(-\omega;\omega,0,0)\epsilon_{\omega j}\epsilon_{0 k}\epsilon_{0 l} cos(\omega t)\nonumber\\
+\frac{1}{8}\gamma^{(2)}_{ijkl}(-\omega;\omega,-\omega,\omega)\epsilon_{\omega j}\epsilon_{\omega k}\epsilon_{\omega\rho} cos(\omega t)\nonumber\\
+\frac{1}{4}\gamma^{(2)}_{ijkl}(-2\omega;\omega,\omega,0)\epsilon_{\omega\mu}\epsilon_{\omega\nu}\epsilon_{0\rho} cos(2\omega t)\nonumber\\
+\frac{1}{4}\gamma^{(2)}_{ijkl}(0;\omega,-\omega,0)\epsilon_{\omega\mu}\epsilon_{\omega\nu}\epsilon_{0\rho} cos(2\omega t)\nonumber\\
+\frac{1}{24}\gamma^{(2)}_{ijkl}(-3\omega;\omega,\omega,\omega)\epsilon_{\omega\mu}\epsilon_{\omega\nu}\epsilon_{\omega\rho} cos(3\omega t)\nonumber\\
\end{eqnarray}

Coupled perturbed Hartree-Fock method implemented in $gaussian09$ 
is used to calculate frequency independent and
frequency dependent linear and
non-linear coefficients. 
Static second order hyper polarizability $\gamma^{(2)}(0;0,0,0)$ and
frequency dependent dc-Kerr coefficients 
($\gamma^{(2)}(-\omega;\omega,0,0)$), the electric field induced
second harmonic (EFISH) ($\gamma^{(2)}(-2\omega;\omega,0,0)$) 
and degenerate four wave mixing(DFWM), 
$\gamma^{(2)}(-\omega;\omega,-\omega,\omega)$
are calculated. 
In case of Dc-Kerr one measures $\gamma_{k}^{(2)}=\frac{3}{2}(\gamma_{\parallel}-\gamma_{\perp})$\cite{shelton1994measurements}, where
\begin{equation}
\gamma_{\parallel}=\frac{1}{15}\sum\limits_{\xi\eta}\{\gamma_{\xi\xi\eta\eta}+\gamma_{\xi\eta\eta\xi}+\gamma_{\xi\eta\xi\eta}\}
\end{equation}
and
\begin{equation}
\gamma_{\perp}=\frac{1}{15}\sum\limits_{\xi\eta}\{2\gamma_{\xi\eta\eta\xi}-\gamma_{\xi\xi\eta\eta}\}
\end{equation}
where $\eta,\xi=x,y,z$ and $\gamma_{\parallel}$ 
is the second hyper polarizability when the field is parallely polarized
and $\gamma_{\perp}$ is the second 
hyper polarizability when the field is perpendicularly polarized. 
\begin{eqnarray}
\gamma_{DFWM}^{(2)}(-\omega;\omega,-\omega,\omega)\approx\frac{1}{3}\gamma_{k}^{(2)}(-2\omega;\omega,\omega,0)+\gamma_{k}^{(2)}(-\omega;\omega,0,0)\nonumber\\-\frac{1}{3}\gamma_{k}^{(2)}(0;0,0,0)
\end{eqnarray}
as $\gamma_{k}^{(2)}=\frac{3}{2}(\gamma_{\parallel}-\gamma_{\perp})$ equation 8 can be written as:
\begin{eqnarray}
\gamma_{DFWM}^{(2)}(-\omega;\omega,-\omega,\omega)\approx\frac{1}{2}\{\gamma_{\parallel}^{(2)}(-2\omega;\omega,\omega,0)-\gamma_{\perp}^{(2)}(-2\omega;\omega,\omega,0)\}\nonumber\\+\frac{3}{2}\{\gamma_{\parallel}^{(2)}(-\omega;\omega,0,0)-\gamma_{\perp}^{(2)}(-\omega;\omega,0,0)\}\nonumber\\-\frac{1}{2}\{\gamma_{\parallel}^{(2)}(0;0,0,0)-\gamma_{\perp}^{(2)}(0;0,0,0)\}
\end{eqnarray}
Finally from DFWM properties quadratic non-linear refractive index can be calculated using the formula:
\begin{equation}
n_{2}\left(\frac{cm^{2}}{W}\right)=8.28\times10^{-23}\gamma_{DFWM}^{(2)}(-\omega;\omega,-\omega,\omega) (a.u.).
\end{equation}
\section{RESULTS AND DISCUSSION:}
\subsection{TD-DFT results for expanded porphyrin systems}
\begin{figure}[!htbp]
\begin{center}
\includegraphics[width=0.50\linewidth]{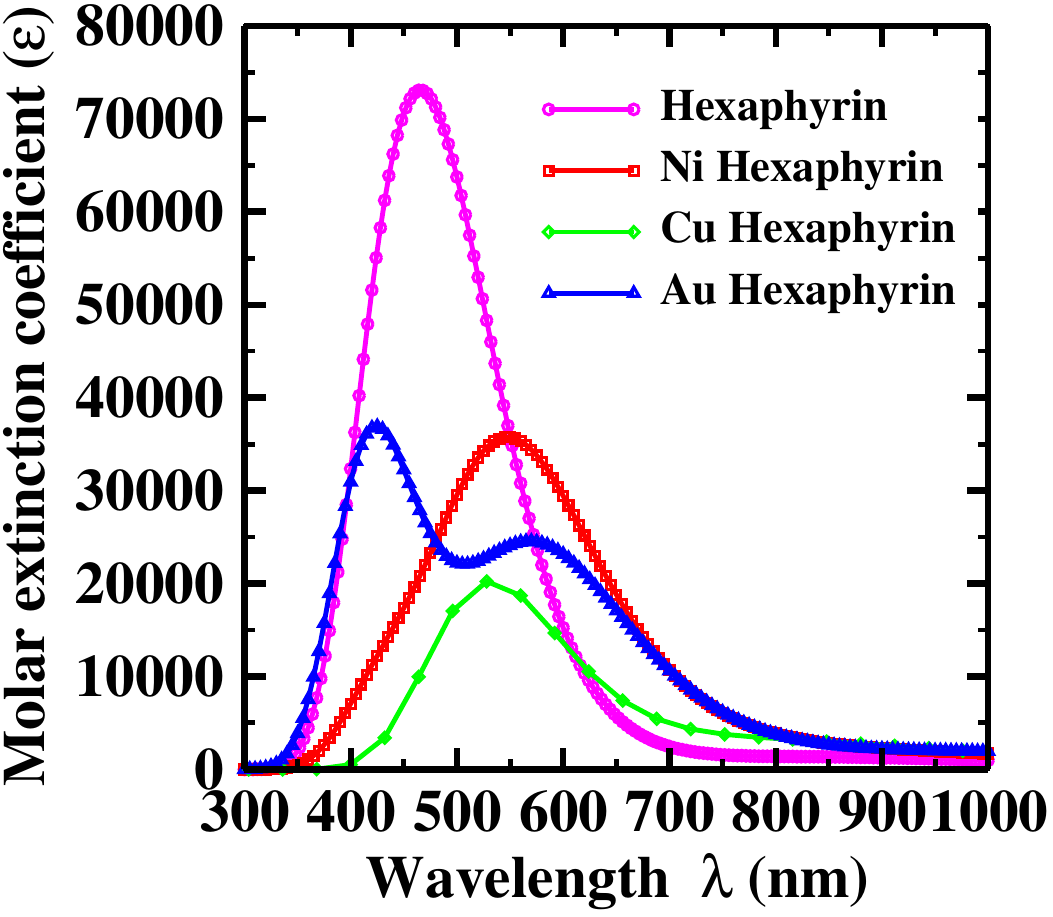}
\caption{Computed UV-vis spectra for hexaphyrin systems}
\label{UV-vis}
\end{center}
\end{figure}

\begin{figure}[!htbp]
\begin{center}
\includegraphics[width=0.80\linewidth]{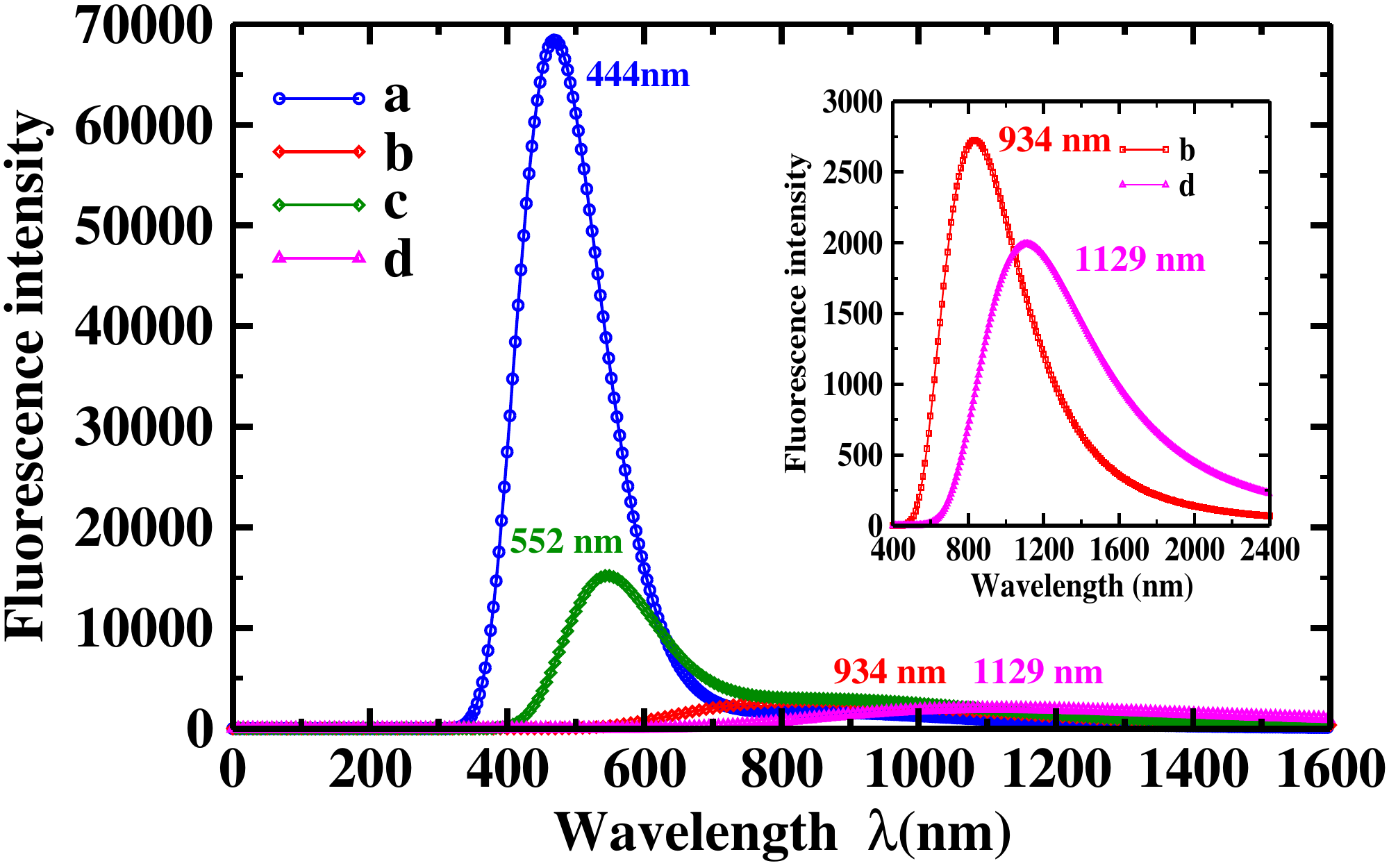}
\caption{Computed emission spectra for hexaphyrin systems}
\label{fluoro}
\end{center}
\end{figure}

The optical properties of these systems are calculated theoretically(Fig:\ref{UV-vis}) has an agreement with the experimental results
\ref{table1}. Use of metal center increases the charge transfer process \cite{ref22} thus the $\Delta_{ST}$ gap can be
theoretically tune to exhibit various optical processes.Expanded monomer with carbazole as donor and -COOH as acceptor shows maximum absorption peak near 501 nm which matches the experimental
value for normal porphyrin with same D-A \cite{ref7}. Use of D-porphyrin-COOH network minimizes the recombination procedure and helps to transfer photo generated charge carriers\cite{ref30}. TDDFT calculation shows broad absorption bands of these molecules
which can be used to harvest solar energy in visible and infrared region. Small $\Delta_{ST}$ gap is useful for reverse
intersystem crossing as it was found in our previous calculation\cite{sumit-mousumi} and it is useful to the fabrication
of organic light emitting diodes (OLED) which demands RISC. As the energy gap $\Delta_{ST}$ computed between 0.7 eV to
0.91 eV the singlet state can be repopulated by the triplets thermally. Experimental values for maximum absorption  as in 
case of hexaphyrin shows peaks at 568 nm, 720 nm, 900 nm and 1018 nm\cite{ref10,ref12,ahn2005comparative} where the experiment was done in dicholoromethane (DCM) solution and the end group is substituted with fluorobenzene. For $\bf b$ and $\bf c$ the end group substitution 
and solution phase during UV/vis calculation are same. For $\bf b$ the absorption peaks are found at 362 nm, 454 nm, 
521 nm, 577 nm and 684 nm\cite{srinivasan2003doubly} and for $\bf c$ the peaks are at 400 nm, 610 nm, 920 nm and 1080 nm\cite{yamasumi2017copper}. For $\bf d$ the 
solution is still DCM but now the meso positions of the hexaphyrin system are substituted by fluorobenzene and the $\beta$ positions are substituted with phenyl group have absorption peaks at 390 nm, 500 nm, 680 nm, 810 nm and 1210 nm\cite{mori2007peripheral}. All these values
 are in good agreement with our computed values\ref{table1}
 The one photon gap $S_{0}-S_{1}$ for these molecules are higher in energy from 0.0-0.5 eV with respect to $2(S_{0}-T_{1})$.
 The TDDFT calculation shows few values of $S_{0}-S_{1}$ below the values we report with zero oscillator strength, 
thus we neglected them. So, all the molecule
studied here shows satisfactory energy condition which can validate SF, however the second condition is not satisfied
\ref{SF2}. These materials can be used to improve the photo-conversion efficiency of single junction solar cell by 
producing multiple excitons at a time. Thus they are capable of both harvestation of solar spectrum in visible and infrared 
region as well as SF materials to improve the performance of single junction solar cell by crossing SQ limit.\\
The emission spectra calculated in TDDFT method reveals that for molecule a and c there is high value of fluorescence intensity, but for molecule b and d
the intensity is feeble (Fig:\ref{fluoro}). The Stoke's shift for molecule a and c is minimal (5 and 16 nm respectively)
which indicates that the excitation energy is almost completely emitted giving
rise to the high fluorescence intensity(Table:\ref{table1}). For molecule b and the the shift is
high (381 and 566 nm respectively) which implies loss of energy between excitation and emission(Table:\ref{table1}).
\begin{table}[!htbp]
\begin{center}
\caption {TD-DFT results of expanded porphyrin monomer}
\label{table1}
\begin{tabular}{ccccccc}
 \hline 
 \hline
Molecule & $S_1$ & $T_n$ & f & Experimental& Stoke's shift \\ 
       & (eV)   & (eV)   &   & $S_1$ (eV)& (nm)   \\      
\hline
           &1.44  &0.53  &0.031 &1.22 & \\
           &1.63  &0.84  &0.001 &1.38 & \\
$\bf a$    &2.44  &      &0.415 &1.72 &5\\
           &2.54  &      &0.720 &2.18 \cite{ref10,ref12,ahn2005comparative}& \\
           &2.82  &      &0.818 &    & \\
\hline
              &1.32 &0.41 &0.031 &1.81 &\\
              &2.09 &0.57 &0.075 &2.15 &\\
$\bf b$       &2.13 &     &0.105 &2.38 &381\\
              &2.24 &     &0.329 &2.73 &\\
              &2.80 &     &0.161 &3.42\cite{srinivasan2003doubly} &\\
\hline
              &1.44 &0.72 &0.040 &1.15 &\\
              &1.58 &0.82 &0.030 &1.35 &\\
$\bf c$       &1.66 &     &0.006 &2.03 &16\\
              &2.31 &     &0.222 &3.10\cite{yamasumi2017copper} &\\
\hline
              &1.18 &0.48 &0.040 &1.02 &\\
              &2.07 &0.58 &0.263 &1.53 &\\
$\bf d$       &2.20 &     &0.337 &1.82 &566\\
              &3.13 &     &0.031 &2.48 &\\
              &     &     &      &3.18\cite{mori2007peripheral} &\\
\hline
\end{tabular}
\end{center}
\end{table} 
\FloatBarrier
\FloatBarrier
\FloatBarrier
\subsection{Computed NLO properties}
\begin{figure}[!htbp]
\begin{center}
\includegraphics[width=1.0\linewidth]{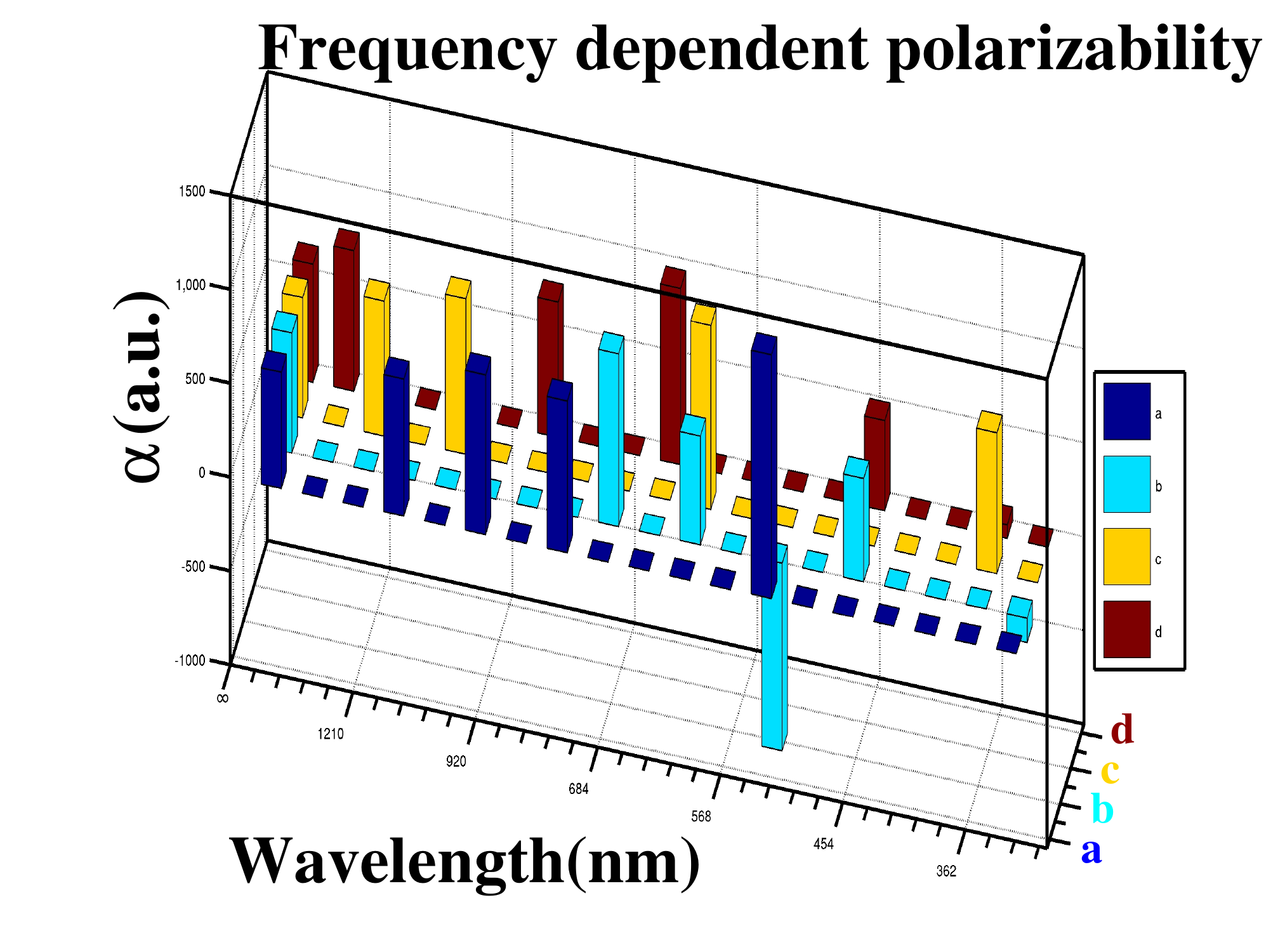}
\label{bar-alpha}
\caption{\bf Frequency dependent polarizability}
\end{center}
\end{figure}

\begin{figure}[!htbp]
\begin{center}
\includegraphics[width=1.0\linewidth]{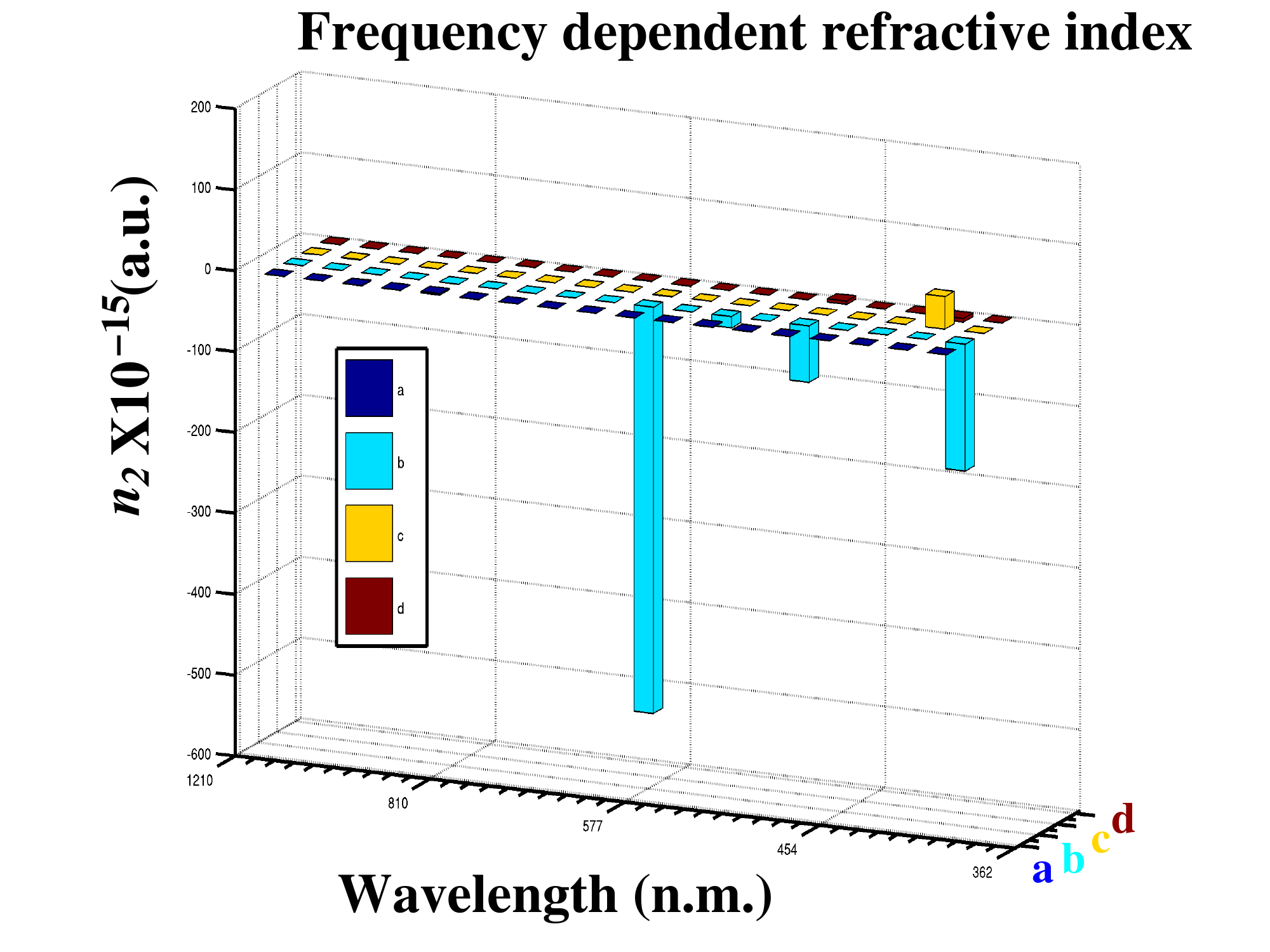}
\label{bar-n2}
\caption{\bf Frequency dependent quadratic refractive index}
\end{center}
\end{figure}

The NLO properties are calculated using coupled perturbed Hartree-Fock method. 
The frequency dependent calculation (done for the frequencies where the experimental absorption peaks are found to be maximum)
 shows extended nature of higher order non linearities. The reason behind choosing the frequency where the maximum
absorption takes place is due reason that at these frequency the phase matching occurs which is the essential criteria
for any NLO phenomenon. It can be shown from our calculation that the $\alpha_{avg}$ values for $\bf a$ increased with
frequency but the nonlinearity coming for the systems $\bf b$, $\bf c$ and $\bf d$ (Fig:\ref{bar-alpha} and table:\ref{table2},table:\ref{table3}). The $\beta$ 
and $\gamma$ values are increased due to the application of core modification\ref{table2},\ref{table3}. The quadratic non 
linear refractive index shows both positive and negative values (Fig:\ref{bar-n2} and table:\ref{table4}) which can be used to modulate LASER frequency 
for both one and two photon absorption.
\FloatBarrier
\FloatBarrier
\begin{table}[!htbp]
\begin{center}
\caption{Frequency independent average polarizability $\alpha_{avg}$, first and second order hyper polarizability $\beta$ and $\gamma$}
\label{table2}
\begin{tabular}{cccccc}
\hline
\hline
Molecule &$\alpha_{avg}(0;0)$ & $\beta_{\parallel}(0;0,0)$ & $\beta_{\perp}(0;0,0)$ & $\gamma_{\parallel}(0;0,0,0)$ &$\gamma_{\perp}(0;0,0,0)$\\
       &(a.u.)  & (a.u)                      & (a.u)                  & (a.u)                         & (a.u) \\
\hline
$\bf a$&0.62$\times 10^3$ &-0.12$\times 10^{-1}$ &-0.40$\times 10^{-2}$ &-0.45$\times 10^5$&-0.15$\times 10^5$ \\
\hline
$\bf b$&0.64$\times 10^3$ &0.18$\times 10^3$ &0.61$\times 10^2$ &-0.58$\times 10^5$ &-0.19$\times 10^5$ \\
\hline
$\bf c$&0.64$\times 10^3$ &0.76$\times 10^{-2}$ &0.25$\times 10^{-2}$ &-0.19$\times 10^6$ &-0.63$\times 10^5$ \\
\hline
$\bf d$&0.63$\times 10^3$ &-0.32$\times 10^0$ &-0.11$\times 10^0$ &-0.24$\times 10^5$ &-0.81$\times 10^4$ \\
\hline
\end{tabular}
\end{center}
\end{table}
\FloatBarrier
\begin{table}[!htbp]
\begin{center}
\caption{Frequency dependent average polarizability $\alpha_{avg}$ and first order hyper polarizability $\beta$}
\label{table3}
\begin{tabular}{ccccccc}
\hline
\hline
Molecule& $\omega$ &$\alpha_{avg}(-\omega;\omega)$ &$\beta_{\parallel}(-2\omega;\omega,\omega)$ & $\beta_{\perp}(-2\omega;\omega,\omega)$  & $\beta_{\parallel}(-\omega;\omega,0)$&$\beta_{\perp}(-\omega;\omega,0)$ \\
     & (a.u.)& (a.u.)     & (a.u.)       & (a.u.)   & (a.u.)  & (a.u.) \\
\hline
$\bf a$   &0.045&0.73$\times 10^3$ &0.10$\times 10^1$ &-0.44$\times 10^0$ &-0.42$\times 10^0$ &-0.42$\times 10^0$\\
          &0.050&0.85$\times 10^3$ &-0.16$\times 10^1$ &0.93$\times 10^1$ &-0.20$\times 10^1$ &-0.23$\times 10^1$\\
          &0.063&0.81$\times 10^3$ &-0.55$\times 10^1$ &0.47$\times 10^1$ &0.64$\times 10^0$ &0.48$\times 10^0$ \\
          &0.080&0.13$\times 10^4$ &-0.48$\times 10^1$ &0.10$\times 10^2$ &0.13$\times 10^1$ &0.12$\times 10^1$  \\
\hline
$\bf b$   &0.066&0.92$\times 10^3$ &0.36$\times 10^4$ &-0.46$\times 10^4$ &-0.39$\times 10^3$ &-0.43$\times 10^3$\\
          &0.079&0.58$\times 10^3$ &0.26$\times 10^6$ &0.18$\times 10^7$ &0.20$\times 10^5$ &0.31$\times 10^5$\\
          &0.087&-0.10$\times 10^4$ &-0.25$\times 10^5$ &-0.18$\times 10^5$ &0.38$\times 10^5$ &0.78$\times 10^5$ \\
          &0.100&0.55$\times 10^3$ &-0.16$\times 10^7$ &0.12$\times 10^6$ &0.22$\times 10^6$ &0.10$\times 10^6$  \\
          &0.126&-0.13$\times 10^3$ &-0.35$\times 10^6$ &-0.65$\times 10^6$ &-0.65$\times 10^5$ &-0.12$\times 10^6$  \\
\hline
$\bf c$   &0.042&0.72$\times 10^3$ &0.66$\times 10^{-1}$ &0.38$\times 10^0$ &-0.12$\times 10^{-1}$ &0.10$\times 10^{-2}$\\
          &0.049&0.83$\times 10^3$ &0.27$\times 10^0$ &-0.86$\times 10^{-2}$ &-0.84$\times 10^{-1}$ &-0.11$\times 10^{-1}$\\
          &0.075&0.98$\times 10^3$ &0.16$\times 10^1$ &0.70$\times 10^0$ &-0.12$\times 10^0$ &-0.10$\times 10^0$ \\
          &0.114&0.75$\times 10^3$ &0.49$\times 10^0$ &0.31$\times 10^1$ &-0.13$\times 10^2$ &-0.42$\times 10^1$  \\
\hline
$\bf d$   &0.037&0.75$\times 10^3$ &0.13$\times 10^1$ &0.19$\times 10^1$ &-0.17$\times 10^1$ &-0.18$\times 10^1$\\
          &0.056&0.72$\times 10^3$ &0.73$\times 10^1$ &0.10$\times 10^2$ &0.95$\times 10^0$ &0.75$\times 10^0$\\
          &0.067&0.94$\times 10^3$ &0.18$\times 10^0$ &0.60$\times 10^0$ &0.12$\times 10^1$ &0.12$\times 10^1$ \\
          &0.091&0.48$\times 10^3$ &-0.70$\times 10^2$ &-0.16$\times 10^3$ &-0.22$\times 10^1$ &-0.33$\times 10^1$  \\
          &0.116&0.72$\times 10^2$ &-0.93$\times 10^2$ &0.74$\times 10^2$ &0.82$\times 10^1$ &-0.23$\times 10^2$  \\
\hline
\end{tabular}
\end{center}
\end{table}
\FloatBarrier
\begin{table}[!htbp]
\begin{center}
\caption{Frequency dependent average second order hyper polarizability $\gamma$}
\label{table4}
\begin{tabular}{cccccc}
\hline
\hline
Molecule & $\omega$ &$\gamma_{\parallel}(-2\omega;\omega,\omega,0)$ & $\gamma_{\perp}(-2\omega;\omega,\omega,0)$  & $\gamma_{\parallel}(-\omega;\omega,0,0)$&$\gamma_{\perp}(-\omega;\omega,0,0)$ \\
     & (a.u.)     & (a.u.)       & (a.u.)   & (a.u.)  & (a.u.) \\ 
\hline
Hexaphyrin &0.045 &-0.52$\times 10^6$ &0.75$\times 10^7$ &0.18$\times 10^6$ & 0.74$\times 10^5$\\
           &0.050 &0.77$\times 10^8$ &0.44$\times 10^8$ &0.22$\times 10^7$ &0.75$\times 10^6$\\
           &0.063 &0.11$\times 10^7$ &-0.82$\times 10^6$ &-0.38$\times 10^6$ &-0.63$\times 10^5$ \\
           &0.080 &-0.13$\times 10^8$ &-0.85$\times 10^7$ &-0.15$\times 10^7$ &-0.61$\times 10^5$  \\     
\hline
Ni Hexaphyrin &0.066 &0.36$\times 10^6$ &-0.44$\times 10^6$ &-0.66$\times 10^6$ & -0.22$\times 10^6$\\
           &0.079 &-0.13$\times 10^{11}$ &-0.82$\times 10^9$ &-0.45$\times 10^7$ &-0.13$\times 10^8$\\
           &0.087 &-0.24$\times 10^9$ &0.87$\times 10^9$ &0.35$\times 10^9$ &0.92$\times 10^8$ \\
           &0.100 &-0.11$\times 10^{10}$ &-0.25$\times 10^8$ &-0.44$\times 10^8$ &0.16$\times 10^9$  \\
           &0.126 &0.34$\times 10^9$ &-0.45$\times 10^9$ &0.15$\times 10^{10}$ &0.48$\times 10^9$  \\
\hline
Cu Hexaphyrin &0.042 &-0.58$\times 10^7$ &0.11$\times 10^7$ &0.22$\times 10^6$ & 0.16$\times 10^6$\\
           &0.049 &-0.47$\times 10^6$ &-0.32$\times 10^6$ &0.11$\times 10^7$ &0.32$\times 10^6$\\
           &0.075 &0.34$\times 10^6$ &0.13$\times 10^6$ &-0.38$\times 10^7$ &-0.12$\times 10^7$ \\
           &0.114 &-0.11$\times 10^9$ &-0.24$\times 10^8$ &0.42$\times 10^9$ &0.70$\times 10^8$  \\
\hline
Au Hexaphyrin &0.037 &-0.59$\times 10^7$ &0.33$\times 10^7$ &0.65$\times 10^6$ & 0.21$\times 10^6$\\
           &0.056 &0.11$\times 10^8$ &0.57$\times 10^7$ &-0.20$\times 10^6$ &-0.23$\times 10^5$\\
           &0.067 &0.40$\times 10^6$ &-0.40$\times 10^7$ &-0.48$\times 10^6$ &-0.81$\times 10^5$ \\
           &0.091 &0.18$\times 10^9$ &0.74$\times 10^8$ &-0.26$\times 10^7$ &-0.74$\times 10^6$  \\
           &0.116 &0.30$\times 10^9$ &0.12$\times 10^9$ &0.36$\times 10^8$ &0.13$\times 10^8$  \\
\hline
\end{tabular}
\end{center}
\end{table}
\FloatBarrier
\begin{table}[!htbp]
\caption{ Values for $\gamma_{DFWM}$ and $n_{2}\left(\frac{cm^{2}}{W}\right)$}
\begin{center}
\begin{tabular}{cccc}
\hline
Molecule & $\omega$ & $\gamma_{DFWM}$ & $ n_{2}\left(\frac{cm^{2}}{W}\right)$\\
         & (a.u.)   & (a.u.)          & (a.u.) \\
\hline 
\hline
Hexaphyrin  &0.045  &-3.83$\times10^6$   &-3.17$\times10^{-16}$ \\
            &0.050  &1.87$\times10^7$   &1.55$\times10^{-15}$ \\
            &0.063  &4.99$\times10^5$   &4.13$\times10^{-17}$ \\
            &0.080  &-4.39$\times10^6$   &-3.63$\times10^{-16}$ \\
\hline
Ni Hexaphyrin &0.066   &-2.40$\times10^5$   &-1.98$\times10^{-17}$ \\
              &0.079   &-6.07$\times10^9$   &-5.02$\times10^{-13}$ \\
              &0.087   &-1.68$\times10^8$   &-1.39$\times10^{-14}$ \\
              &0.100   &-8.43$\times10^8$   &-6.98$\times10^{-14}$ \\
              &0.126   &1.90$\times10^9$   &1.57$\times10^{-13}$ \\
\hline
Cu Hexaphyrin &0.042   &-3.29$\times10^6$  &-2.72$\times10^{-16}$ \\
              &0.049   &1.16$\times10^6$  &9.60$\times10^{-17}$ \\
              &0.075   &-3.73$\times10^6$  &-3.09$\times10^{-16}$ \\
              &0.114   &4.85$\times10^8$  &4.01$\times10^{-14}$ \\
\hline
Au Hexaphyrin &0.037   &-3.93$\times10^6$  &-3.25$\times10^{-16}$ \\
              &0.056   &2.35$\times10^6$  &1.94$\times10^{-16}$ \\
              &0.067   &1.61$\times10^6$  &1.33$\times10^{-16}$ \\
              &0.091   &5.02$\times10^7$  &4.15$\times10^{-15}$ \\
              &0.116   &-5.57$\times10^7$  &-4.61$\times10^{-15}$ \\
\hline
\end{tabular}
\end{center}
\end{table}
\FloatBarrier
The sign of DFWM and $n_2$ depends on the contribution from one photon and two photon. If the two photon contribution is
dominating in the non-resonant regime then the sign of DFWM and $n_2$ will be positive and vice-versa\cite{stegeman2011off}.
\section{CONCLUSION}
We concluded that the range of the absorption band of EPs are diverse in nature. Thus solar energy can be harvested in 
visible as well as near infrared region of the solar spectrum. Here we got the values for singlet to triplet gap ($\Delta E_{ST}$) are ranging from 0.70 to 0.91 which is quite small than the normal porphyrin dimer, for normal porphyrin these values 
ranges from 1.52-1.69\cite{ref25}. Thus EPs are showing better results for $\Delta E_{ST}$ compared to porphyrin. When we 
use metal center the absorption peaks are quite similar to the experimental result of EPs whose core modified with metal 
center. All the systems studied satisfy the first condition for SF, which can be applicable to design potential SF material
to overcome theoretical SQ limit. Small $\Delta_{ST}$ gap helps the triplet excitons to populate themselves to singlet
exciton states through RISC and made possible for the fabrication of OLED. The emission spectra shows for molecule a and c
the Stoke's shift is minimum giving rise to higher fluorescence intensity compared to molecule b and d with large Stoke's shift. The NLO properties studied shows higher order
nonlinearity in hyperpolarizabilty. These properties have many applications in the field of electro-optic measurements,
sensing, LASERS and many more. The quadratic non linear refractive index $n_2$ shows both positive and negative values
which indicates that for both two photon and one photon frequency the modulation of LASER intensity can be done. So we can 
choose the porphyrin systems for our opto-electronic and NLO calculations as these systems shows diverse application. 
Successful tuning of the $\Delta_{ST}$ gap and frequency lead these materials to fabricate electro-optic miniaturizer.
\section{ACKNOWLEDGMENT}
Authors acknowledge financial support from Department of Science and Technology,
Government of India through a SERB Fast-Track Grant
SB/FT/CS-164/2013 and IISER Kolkata for fellowship. We also thankful to Mr. Kingsuk Mukhuti for fruitful scientific and technical discussion.
\bibliographystyle{unsrt}
\bibliography{NLO}

\begin{thebibliography}{10}

\bibitem{smith2010singlet}
Millicent~B Smith and Josef Michl.
\newblock Singlet fission.
\newblock {\em Chemical reviews}, 110(11):6891--6936, 2010.

\bibitem{naskar2020use}
Sumit Naskar and Mousumi Das.
\newblock The use of low-lying excited states of zethrene and its homologs in
  singlet fission within pariser-parr-pople model hamiltonian: A density matrix
  renormalization group study.
\newblock {\em Chemical Physics}, page 110717, 2020.

\bibitem{shockley1961detailed}
William Shockley and Hans~J Queisser.
\newblock Detailed balance limit of efficiency of p-n junction solar cells.
\newblock {\em Journal of applied physics}, 32(3):510--519, 1961.

\bibitem{paci2006SF}
Irina Paci, Justin~C Johnson, Xudong Chen, Geeta Rana, Du{\v{s}}ka Popovi{\'c},
  Donald~E David, Arthur~J Nozik, Mark~A Ratner, and Josef Michl.
\newblock Singlet fission for dye-sensitized solar cells: Can a suitable
  sensitizer be found?
\newblock {\em Journal of the American Chemical Society}, 128(51):16546--16553,
  2006.

\bibitem{oudar-chemla}
J-L\_ Oudar and DS~Chemla.
\newblock Hyperpolarizabilities of the nitroanilines and their relations to the
  excited state dipole moment.
\newblock {\em The Journal of Chemical Physics}, 66(6):2664--2668, 1977.

\bibitem{marder}
Seth~R Marder.
\newblock Organic nonlinear optical materials: where we have been and where we
  are going.
\newblock {\em Chemical communications}, (2):131--134, 2006.

\bibitem{bache2013anisotropic}
Morten Bache, Hairun Guo, Binbin Zhou, and Xianglong Zeng.
\newblock The anisotropic kerr nonlinear refractive index of the beta-barium
  borate ($\beta$-bab 2 o 4) nonlinear crystal.
\newblock {\em Optical Materials Express}, 3(3):357--382, 2013.

\bibitem{makhal2016third}
Krishnandu Makhal, Shafali Arora, Paramjit Kaur, Debabrata Goswami, and
  Kamaljit Singh.
\newblock Third-order nonlinear optical response and ultrafast dynamics of
  tetraoxa [22] porphyrin (2.1. 2.1) s.
\newblock {\em Journal of Materials Chemistry C}, 4(40):9445--9453, 2016.

\bibitem{karakas2019theoretical}
A~Karakas, Y~Ceylan, M~Karakaya, M~Taser, BB~Terlemez, N~Eren, Y~El~Kouari,
  M~Lougdali, AK~Arof, and B~Sahraoui.
\newblock Theoretical diagnostics of second and third-order
  hyperpolarizabilities of several acid derivatives.
\newblock {\em Open Chemistry}, 17(1):151--156, 2019.

\bibitem{nalwa-jpcA-1993}
Hari~Singh Nalwa, Atsushi Kakuta, and Akio Mukoh.
\newblock Third-order nonlinear optical properties of a vanadyl
  naphthalocyanine derivative.
\newblock {\em The Journal of Physical Chemistry}, 97(6):1097--1100, 1993.

\bibitem{drobizhev-cpl-2002}
M~Drobizhev, A~Karotki, M~Kruk, N~Zh Mamardashvili, and A~Rebane.
\newblock Drastic enhancement of two-photon absorption in porphyrins associated
  with symmetrical electron-accepting substitution.
\newblock {\em Chemical physics letters}, 361(5-6):504--512, 2002.

\bibitem{simil-JPCA-2013}
Simil Thomas, YA~Pati, and S~Ramasesha.
\newblock Linear and nonlinear optical properties of expanded porphyrins: A
  dmrg study.
\newblock {\em The Journal of Physical Chemistry A}, 117(33):7804--7809, 2013.

\bibitem{ganeev2003nonlinear}
RA~Ganeev, IA~Kulagin, AI~Ryasnyanskii, RI~Tugushev, and T~Usmanov.
\newblock The nonlinear refractive indices and nonlinear third-order
  susceptibilities of quadratic crystals.
\newblock {\em Optics and Spectroscopy}, 94(4):561--568, 2003.

\bibitem{ref16}
M.~J. Frisch, G.~W. Trucks, H.~B. Schlegel, G.~E. Scuseria, M.~A. Robb, J.~R.
  Cheeseman, G.~Scalmani, V.~Barone, B.~Mennucci, G.~A. Petersson,
  H.~Nakatsuji, M.~Caricato, X.~Li, H.~P. Hratchian, A.~F. Izmaylov, J.~Bloino,
  G.~Zheng, J.~L. Sonnenberg, M.~Hada, M.~Ehara, K.~Toyota, R.~Fukuda,
  J.~Hasegawa, M.~Ishida, T.~Nakajima, Y.~Honda, O.~Kitao, H.~Nakai, T.~Vreven,
  J.~A. Montgomery, J.~E.~Peralta Jr., F.~Ogliaro, M.~Bearpark, J.~J. Heyd,
  E.~Brothers, K.~N. Kudin, V.~N. Staroverov, R.~Kobayashi, J.~Normand,
  K.~Raghavachari, A.~Rendell, J.~C. Burant, S.~S. Iyengar, J.~Tomasi,
  M.~Cossi, N.~Rega, J.~M. Millam, M.~Klene, J.~E. Knox, J.~B. Cross,
  V.~Bakken, C.~Adamo, J.~Jaramillo, R.~Gomperts, R.~E. Stratmann, O.~Yazyev,
  A.~J. Austin, R.~Cammi, C.~Pomelli, J.~W. Ochterski, R.~L. Martin,
  K.~Morokuma, V.~G. Zakrzewski, G.~A. Voth, P.~Salvador, J.~J. Dannenberg,
  S.~Dapprich, A.~D. Daniels, \''O. Farkas, J.~B. Foresman, J.~V. Ortiz,
  J.~Cioslowski, and D.~J. Fox.
\newblock G09 gaussian inc.
\newblock {\em Wallingford, CT}, 2009.

\bibitem{ref18}
Axel~D Becke.
\newblock Density-functional exchange-energy approximation with correct
  asymptotic behavior.
\newblock {\em Physical review A}, 38(6):3098, 1988.

\bibitem{ref19}
Chengteh Lee, Weitao Yang, and Robert~G Parr.
\newblock Development of the colle-salvetti correlation-energy formula into a
  functional of the electron density.
\newblock {\em Physical review B}, 37(2):785, 1988.

\bibitem{ref17}
Erich Runge and Eberhard~KU Gross.
\newblock Density-functional theory for time-dependent systems.
\newblock {\em Physical Review Letters}, 52(12):997, 1984.

\bibitem{tarazkar2015theoretical}
Maryam Tarazkar, Dmitri~A Romanov, and Robert~J Levis.
\newblock Theoretical study of second-order hyperpolarizability for nitrogen
  radical cation.
\newblock {\em Journal of Physics B: Atomic, Molecular and Optical Physics},
  48(9):094019, 2015.

\bibitem{shelton1994measurements}
David~P Shelton and Julia~E Rice.
\newblock Measurements and calculations of the hyperpolarizabilities of atoms
  and small molecules in the gas phase.
\newblock {\em Chemical Reviews}, 94(1):3--29, 1994.

\bibitem{ref22}
Xing-Yu Li, Cai-Rong Zhang, You-Zhi Wu, Hai-Min Zhang, Wei Wang, Li-Hua Yuan,
  Hua Yang, Zi-Jiang Liu, and Hong-Shan Chen.
\newblock The role of porphyrin-free-base in the electronic structures and
  related properties of n-fused carbazole-zinc porphyrin dye sensitizers.
\newblock {\em International journal of molecular sciences},
  16(11):27707--27720, 2015.

\bibitem{ref7}
Shohei Saito and Atsuhiro Osuka.
\newblock Expanded porphyrins: intriguing structures, electronic properties,
  and reactivities.
\newblock {\em Angewandte Chemie International Edition}, 50(19):4342--4373,
  2011.

\bibitem{ref30}
Aswani Yella, Hsuan-Wei Lee, Hoi~Nok Tsao, Chenyi Yi, Aravind~Kumar Chandiran,
  Md~Khaja Nazeeruddin, Eric Wei-Guang Diau, Chen-Yu Yeh, Shaik~M Zakeeruddin,
  and Michael Gr{\"a}tzel.
\newblock Porphyrin-sensitized solar cells with cobalt (ii/iii)--based redox
  electrolyte exceed 12 percent efficiency.
\newblock {\em science}, 334(6056):629--634, 2011.

\bibitem{sumit-mousumi}
Sumit Naskar and Mousumi Das.
\newblock Singlet and triplet excited state energy ordering in cyclopenta-fused
  polycyclic aromatic hydrocarbons (cp-pahs) suitable for energy harvesting: An
  exact model and tddft study.
\newblock {\em ACS Omega}, 2(5):1795--1803, 2017.

\bibitem{ref10}
Hirotaka Mori, Takayuki Tanaka, Sangsu Lee, Jong~Min Lim, Dongho Kim, and
  Atsuhiro Osuka.
\newblock meso--meso linked porphyrin--[26] hexaphyrin--porphyrin hybrid arrays
  and their triply linked tapes exhibiting strong absorption bands in the nir
  region.
\newblock {\em Journal of the American Chemical Society}, 137(5):2097--2106,
  2015.

\bibitem{ref12}
Tomoki Yoneda, Taeyeon Kim, Takanori Soya, Saburo Neya, Juwon Oh, Dongho Kim,
  and Atsuhiro Osuka.
\newblock Conformational fixation of a rectangular antiaromatic [28] hexaphyrin
  using rationally installed peripheral straps.
\newblock {\em Chemistry--A European Journal}, 22(13):4413--4417, 2016.

\bibitem{ahn2005comparative}
Tae~Kyu Ahn, Jung~Ho Kwon, Deok~Yun Kim, Dae~Won Cho, Dae~Hong Jeong,
  Seong~Keun Kim, Masaaki Suzuki, Soji Shimizu, Atsuhiro Osuka, and Dongho Kim.
\newblock Comparative photophysics of [26]-and [28] hexaphyrins (1.1. 1.1.
  1.1): Large two-photon absorption cross section of aromatic [26] hexaphyrins
  (1.1. 1.1. 1.1).
\newblock {\em Journal of the American Chemical Society}, 127(37):12856--12861,
  2005.

\bibitem{srinivasan2003doubly}
Alagar Srinivasan, Tomoya Ishizuka, Atsuhiro Osuka, and Hiroyuki Furuta.
\newblock Doubly n-confused hexaphyrin: a novel aromatic expanded porphyrin
  that complexes bis-metals in the core.
\newblock {\em Journal of the American Chemical Society}, 125(4):878--879,
  2003.

\bibitem{yamasumi2017copper}
Kazuhisa Yamasumi, Keiichi Nishimura, Yutaka Hisamune, Yusuke Nagae, Tomoki
  Uchiyama, Kazutaka Kamitani, Tomoyasu Hirai, Maiko Nishibori, Shigeki Mori,
  Satoru Karasawa, et~al.
\newblock Bis-copper (ii)/$\pi$-radical multi-heterospin system with
  non-innocent doubly n-confused dioxohexaphyrin (1.1. 1.1. 1.0) ligand.
\newblock {\em Chemistry--A European Journal}, 23(61):15322--15326, 2017.

\bibitem{mori2007peripheral}
Shigeki Mori, Kil~Suk Kim, Zin~Seok Yoon, Su~Bum Noh, Dongho Kim, and Atsuhiro
  Osuka.
\newblock Peripheral fabrications of a bis-gold (iii) complex of [26]
  hexaphyrin (1.1. 1.1. 1.1) and aromatic versus antiaromatic effect on
  two-photon absorption cross section.
\newblock {\em Journal of the American Chemical Society}, 129(37):11344--11345,
  2007.

\bibitem{stegeman2011off}
George Stegeman, Mark~G Kuzyk, Dimitris~G Papazoglou, and Stelios Tzortzakis.
\newblock Off-resonance and non-resonant dispersion of kerr nonlinearity for
  symmetric molecules.
\newblock {\em Optics express}, 19(23):22486--22495, 2011.

\bibitem{ref25}
A~Suvitha, H~Mizuseki, Y~Kawazoe, M~Takeda, M~Kohno, and N~Ohuchi.
\newblock Td-dft studies on hematoporphyrin and its dimers.
\newblock {\em Materials transactions}, 49(11):2416--2419, 2008.

\end{thebibliography}
\end{document}